# Policy documents as sources for measuring societal impact:

# How often is climate change research mentioned in policy-related documents?

Lutz Bornmann*, Robin Haunschild**, and Werner Marx**

*Division for Science and Innovation Studies

Administrative Headquarters of the Max Planck Society

Hofgartenstr. 8,

80539 Munich, Germany.

E-mail: bornmann@gv.mpg.de

**Max Planck Institute for Solid State Research

Heisenbergstr. 1,

70569 Stuttgart, Germany.

**Abstract**

In the current UK Research Excellence Framework (REF) and the Excellence in Research for Australia (ERA) societal impact measurements are inherent parts of the national evaluation systems. In this study, we deal with a relatively new form of societal impact measurements. Recently, Altmetric – a start-up providing publication level metrics – started to make data for publications available which have been mentioned in policy documents. We regard this data source as an interesting possibility to specifically measure the (societal) impact of research. Using a comprehensive dataset with publications on climate change as an example, we study the usefulness of the new data source for impact measurement. Only 1.2% (n=2,341) out of 191,276 publications on climate change in the dataset have at least one policy mention. We further reveal that papers published in *Nature* and *Science* as well as from the areas "Earth and related environmental sciences" and "Social and economic geography" are especially relevant in the policy context. Given the low coverage of the climate change literature in policy documents, this study can be only a first attempt to study this new source of altmetric data. Further empirical studies are necessary in upcoming years, because mentions in policy documents are of special interest in the use of altmetric data for measuring target-oriented the broader impact of research.

**Key words**

Climate change; Policy documents; altmetrics

# 1     Introduction

Academic science emerged at the beginning of the 19$^{th}$ century (Ziman, 1996). During academic science, the evaluation of scientific results focuses on their excellence and originality in a self-regulated process – the peer review process (Bornmann, 2011; Petit, 2004). The post-academic science which begins in the 1980s (Ziman, 2000) is characterized by an increasing competition for research funds (national and international) which are mostly project-dependent distributed. Applicants of project proposals are more and more forced to be specific about the expected outcome and its wider economic and societal impact (Ziman, 1998): The context of application becomes the interesting topic which decides on funding (besides excellence and originality). The objective of post-academic science "is not scientific excellence and theory-building as such but rather the production of a result that is relevant and applicable for the users of the research; in other words, the result should be socially relevant, socially robust and innovative" (Erno-Kjolhede & Hansson, 2011, p. 134). The new objective also leads to changes in the ex-post evaluation of science. During the 1990s researchers started to develop evaluation systems with possible indicators in order to measure the societal impact of research (Miettinen, Tuunainen, & Esko, 2015). Also, governments and intermediaries (government-funded granting agencies) started requiring that applicants delineate their broader impact plans (Dance, 2013). In the current UK Research Excellence Framework (REF) and the Excellence in Research for Australia (ERA) societal impact measurements are inherent parts of the national evaluation systems.

How is societal impact defined? A pointedly definition has been published by Wilsdon et al. (2015): "Research has a societal impact when auditable or recorded influence is achieved upon non-academic organisation(s) or actor(s) in a sector outside the university sector itself – for instance, by being used by one or more business corporations, government bodies, civil society organisations, media or specialist/professional media organisations or in public debate. As is the case with academic impacts, societal impacts need to be demonstrated rather than assumed. Evidence of external impacts can take the form of references to, citations of or discussion of a person, their work or research results" (p. 6). Samuel and Derrick (2015) interviewed evaluators working in the REF and asked for their definition of societal impact. The interviewees defined societal impact as 'outcome' from research which makes a change or difference (e.g. a change to clinical practice or patient care) or which creates economic benefits (e.g. creating jobs).

In this study we deal with a relatively new form of societal impact measurements. Recently, Altmetric – a start-up providing publication level metrics (www.altmetric.com) – started to make data for academic publications available which have been mentioned in policy documents (Liu, 2014) in order to uncover the interaction between science and politics. On the one side, governments allocate very large amounts of public money to various units (e.g. researchers or institutions) for various forms of research. In many countries, the public money is distributed by the soft money system in which researchers formulate proposals for projects and funding bodies decide on their acceptance or rejection. For governments, academic science is one section of a vague defined Research & Development system which ranges from basic science to near-market technological development (Ziman, 2000). On the other side, independent and still active scientists advice stakeholders in the policy area: "The scientific community is increasingly being called upon to provide evidence and advice to government policy-makers across a range of issues, from short-term public health emergencies through to longer-term challenges, such as population ageing or climate change" (OECD, 2015, p. 5). The scientific advice can be made in direct face-to-face interactions or in indirect interactions whereby papers written by scientists (actually written for their colleagues) are read by political actors and mentioned in policy-related documents. The latter type of interaction can possibly be measured by the new data source offered by Altmetric.

We regard the new data source as an interesting possibility to specifically measure the (societal) impact of research on policy-related areas which should be empirically investigated. Using a comprehensive dataset with publications on climate change as an example, we study the usefulness of the new data source for impact measurement. As publication set in this study, we used climate change literature because (1) corresponding policy sites are continuously analyzed by Altmetric. (2) We expect to observe many references to the scientific literature in the policy documents (Liu, 2014) because climate change is a policy relevant topic since many years. We are especially interested in the characteristics of the papers which are mentioned in the policy documents: Are these papers published in certain journals (e.g. popular journals like *Nature* and *Science*), in certain publication years (e.g. more recent years), or with certain document types (e.g. reviews)?

## 2   Literature overview

Current literature overviews on societal impact studies have been published by Bornmann (2012, 2013). Thus, we concentrate in this chapter on more recently published studies which are especially relevant for this study.

Societal impact measurements are mostly commissioned by governments which argue that measuring the impact on science says little about real-world benefits of research (Cohen et al., 2015). Nightingale and Scott (2007) summarize this argumentation in the following pointedly sentence: "Research that is highly cited or published in top journals may be good for the academic discipline but not for society" (p. 547). Governments are interested to know the importance of public-funded research (1) for the private and public sectors (e.g. health care), (2) to tackle societal challenges (e.g. climate change), and (3) for education and training of the next generations (ERiC, 2010; Grimson, 2014). The impact model of Cleary, Siegfried, Jackson, and Hunt (2013) additionally highlights the policy enactment of research, in which the impact on policies, laws, and regulations is of special interest. The current study seizes upon this additional issue by investigating a possible source for measuring policy enactment of research.

The early studies which targeted societal impact of research focused on the economic dimension. Researchers were interested in the impact of Research and Development (R & D) on economic growth and productivity as well as the rate of return to investments in R & D (Godin & Dore, 2005). The relationships had been quantitatively-statistically investigated. In the current frameworks of national evaluation systems (especially the REF) impact is broader measured (i.e. one focuses not only on the impact of research on economy) and qualitative approaches of measuring impact are rated higher than quantitative approaches. REF evaluation panels review case studies (documents of four pages) which explain how particular research (conducted in the last 15 years and recognized internationally in terms of originality and significance) had influenced the wider, non-academic society (Derrick, Meijer, & van Wijk, 2014; Samuel & Derrick, 2015). According to Miettinen et al. (2015) "in the recent Research Excellence Framework, 6,975 impact case studies were conducted and evaluated by more than 1,000 assessment panel members from academic and societal interest groups".

The use of the case study approach for measuring societal impact is positively as well as negatively assessed in the scientometric literature. The preparation of case studies means a high effort for an institution. David Price, the vice-provost for research at the University College London, said that the university alone "wrote 300 case studies that took around 15 person-years of work, and hired four full-time staff members to help" (Van Noorden, 2015, p. 150). According to Wilsdon et al. (2015) case studies "offer the potential to present complex information and warn against more focus on quantitative metrics for impact case studies. Others however see case studies as 'fairy tales of influence' and argue for a more consistent toolkit of impact metrics that can be more easily compared across and between cases" (p. 49, see also Atkinson, 2014). For Cohen et al. (2015) "the holy grail is to find short term indicators that can be measured before, during or immediately after the research is completed and that are robust predictors of the longer term impact … from the research". Ovseiko, Oancea, and Buchan (2012) think that important areas of impact can only be captured qualitatively. "However, the emphasis on qualitative indicators would stretch traditional peer review further and concentrate on the most prominent examples of impact, overlooking more modest contributions".

Although the literature overview on societal impact studies of Bornmann (2012, 2013) reveal many proposals to measure societal impact quantitatively, Wilsdon et al. (2015) highlight only two metrics which might reflect wider societal impact: (1) Google patent citations and (2) clinical guideline citations. (Ad 1) The analysis of the reference list in patents can be used to assess the contribution of publicly-funded research to innovations in industry (Kousha & Thelwall, in press; Ovseiko et al., 2012). (Ad 2) Grant (1999) proposed to use citations to publications in clinical guidelines to "quantify the progress of knowledge from biomedical research into clinical practice" (p. 33). The analysis of 43 UK guidelines show, i.e., that "within the United Kingdom, Edinburgh and Glasgow stood out for their unexpectedly high contributions to the guidelines' scientific base" (Lewison & Sullivan, 2008, p. 1944). The results of Andersen (2013) point out that the quality of clinical guidelines correlates positively with the citation impact of the cited literature. Thus, quality in research and practice seems to be (closely) related.

In the current discussion around the quantitative measurement of societal impact, alternative metrics (altmetrics) play an important role (Bornmann, 2014). Altmetrics count views, downloads, clicks, notes, saves, tweets, shares, likes, recommends, tags, posts, trackbacks, discussions, bookmarks, and comments to measure the impact of scholarly publications. Altmetrics are seen as an interesting possibility to measure quantitatively the broad impact of publications: "Funders also often see tremendous value in the general public understanding of publicly funded research projects and the scientific process. Use of alternative assessment metrics may help support needs in these areas" (NISO Alternative Assessment Metrics Project, 2014).

There are many universal problems with societal impact measurements (Bornmann, 2012, 2013). Milat, Bauman, and Redman (2015) emphasized that "research impacts are complex, non-linear, and unpredictable in nature and there is a propensity to 'count what can be easily measured', rather than measuring what 'counts' in terms of significant, enduring changes". In order to tackle the complexity of societal impact measurements, proposed models of impact measurements tended to be complexly designed. These models might be able to measure impact multi-dimensionally, but do not have the right balance between comprehensiveness and feasibility (Milat et al., 2015). Another problematic part of societal impact measurement is the long time lag before impact manifests in practice. One can expect around 15 years until research evidence is implemented into practice (Balas & Boren, 2000). The US National Research Council (2014) lists the following additional "barriers to assessing research impacts: research can have both positive and negative effects (e.g., the creation of chlorofluorocarbons reduced stratospheric ozone); the adoption of research findings depends on sociocultural factors; transformative innovations often depend on previous research; it is extremely difficult to assess the individual and collective impacts of multiple researchers who are tackling the same problem; and finally, it is difficult to assess the transferability of research findings to other, unintended problems" (p. 75).

Although mentions of papers in policy documents are not able to solve all these problems, they are an interesting source of data to measure the broader impact of research target-oriented (on policy-related areas in society). Also, the data can possibly serve to shed light on the relationship between academic research and policy making. Decision processes in policy making are "deeply permeated" by active scientists (Lasswell, 1971, p. 125). Scientists participate in different positions and roles in the policy process, which are combined with different strategies and goals. Typologies of the positions and roles are described by Weiss (2003, 2006) and Pielke (2007). Pielke (2007) suggests that the role for scientists in political debate is to "help us to understand the associations between different choices and their outcomes" (p. 139).

Weiss (2003, 2006) developed a typology with five positions which an advising scientist can take when addressing uncertainties. "Each position represents an attitude that results from a given level of uncertainty in combination with differences in the perceived necessity to take measures and in the willingness to do so based on associated (societal) costs" (Spruijt, Knol, Petersen, & Lebret, 2016, p. 45). The typology of Pielke (2007) consists of four roles: The *pure scientist* is not interested on the interaction with political actors; she is only oriented towards facts. The *science arbiter* consents to answer specific questions from political actors (which should be factually oriented). The *issue advocate* promotes one solution in the range of options available to political actors. The *honest broker of policy alternatives* tries to explain and/ or to expand the range of options available to political actors.

There is a wide body of literature available investigating the relationship between academic research and policy making, especially concerning climate research. For example, Ford, Knight, and Pearce (2013) assess the 'usability' of climate change research for decision-making. Using a case study of the Canadian International Polar Year, they introduce a novel approach to quantitatively evaluate the usability of climate change research for informing decision-making. Lemos and Morehouse (2005) examine the use of interactive models of research in the regional integrated scientific assessments (RISAS) in the US. They show that highly interactive models increase the possibilities of societal impact. The recent study of Hermann, Pregernig, Hogl, and Bauer (2015) focusses on the expert advices in the field of Austrian climate policy. The authors map the different actors and advisory forms, assess the relevance of scientific knowledge, and identify patterns of science-policy interactions.

## 3   Methods

### 3.1   Analyses of policy documents at Altmetric

Altmetric has developed a text-mining solution (the Altmetric Policy Miner, APM) to discover mentions of publications in policy documents (Liu, 2014; Liu, Konkiel, & Williams, 2015). The following list shows some sources which are analyzed by Altmetric:

- European Food Safety Authority (EFSA)
- GOV.UK - Policy papers, Research & Analysis
- Intergovernmental Panel on Climate Change (IPCC)
- International Committee of the Red Cross (ICRC)
- World Health Organization (WHO)
- International Monetary Fund (IMF)
- Médicins sans Frontières (MSF)
- NICE Evidence
- Oxfam Policy & Practice
- UNESCO
- World Bank

According to Liu et al. (2015) one limitation of the current policy tracking is that the sources are limited to major organizations from North America and Europe. Altmetric plans to curate and add sources, particularly those from regions such as Asia and Africa.

### 3.2   Climate change publication set

In 2015, we have constructed a set of 222,060 papers (articles and reviews only) on climate change (also referred to as climate change papers, CCP, in the following) published between 1980 and 2014 via a sophisticated search method called "interactive query formulation". A detailed explanation of the procedure can be found in Haunschild, Bornmann, and Marx (conditionally accepted for publication). They used the dataset to study (1) the growth of the overall publication output on climate change as well as (2) of the major subfields, and (3) the contributing journals and countries as well as their citation impact. (4) Furthermore, they illustrated the time evolution and relative importance of specific research topics on the basis of a title word analysis.

Our Web of Science (WoS) in-house database – derived from the Science Citation Index Expanded (SCI-E), Social Sciences Citation Index (SSCI), and Arts and Humanities Citation Index (AHCI) provided by Thomson Reuters (Philadelphia, USA) – has DOIs for 149,657 of the 222,060 papers. We tried to obtain the DOIs for the remaining 72,403 papers from CrossRef using the Perl module Bib::CrossRef.[1] The API returned a DOI for each paper. Therefore, we checked if the publication year, journal title and issue agree with the values in our in-house database. For 42.5% (n=30,784) of the cases, the bibliographic values disagreed and the papers were discarded. Using this procedure, we obtained 41,619 additional DOIs from the climate change publication set.

The DOIs (n=191,276, 86.1% of the full publication set) from our in-house database and the DOIs from CrossRef were used to retrieve information about policy mentions from Altmetric via their API[2] on 16 November, 2015. For 1.2% (n=2,341) of the papers, we found at least one policy mention.

Table 1 shows the policy sources which mention climate change papers. The most mentions of climate change papers originate from two organizations: "Food and Agriculture Organization of the United Nations" and "Intergovernmental Panel on Climate Change".

---

[1] See http://search.cpan.org/dist/Bib-CrossRef/lib/Bib/CrossRef.pm
[2] See https://api.altmetric.com/

Table 1. Policy sources which mention climate change papers

| Policy source | Number of policy mentions of climate change papers |
|---|---|
| Food and Agriculture Organization of the United Nations | 966 |
| Intergovernmental Panel on Climate Change | 866 |
| World Bank | 533 |
| Australian Policy Online | 299 |
| UK Government (GOV.UK) | 284 |
| World Health Organization | 124 |
| European Food Safety Authority | 117 |
| Oxfam GB Policy & Practice | 66 |
| UNESCO | 11 |
| International Monetary Fund | 9 |
| The International Fund for Agricultural Development | 8 |
| The Association of the Scientific Medical Societies in Germany | 5 |
| National Institute for Health and Care Excellence | 3 |
| The Royal Society for the Prevention of Accidents | 2 |
| European Centre for Disease Prevention and Control | 1 |
| National Health and Medical Research Council (NHMRC) (Aus) | 1 |

# 4   Results

## 4.1   Climate change papers mentioned in policy documents

Table 2 shows the number of CCP with one or more mentions in policy documents. As the results in the table show, the papers received up to 18 mentions. However, most of the papers (92%) received only one (78.7%) or two (13.3%) mentions. This means that not only few papers are ever mentioned in policy documents; nearly all mentioned papers received only one or two mentions. In the following we will reveal some characteristics of the CCP which have been mentioned in policy documents. Since the number of mentions per CCP is very low ($\bar{x}=1.4$), the following tables in this section do not consider the number of mentions, but analyze the characteristics of the climate change papers mentioned at least once in policy documents (CCP_P). Section 4.2 discusses the few CCP which were mentioned very often in policy documents.

Table 2. Number of climate change papers with at least one mention in policy documents

| Number of mentions in policy documents | Absolute number of documents (CCP_P) | Proportion of documents |
|---|---:|---|
| 1 | 1,842 | 78.7 |
| 2 | 311 | 13.3 |
| 3 | 90 | 3.8 |
| 4 | 39 | 1.7 |
| 5 | 26 | 1.1 |
| 6 | 11 | 0.5 |
| 7 | 7 | 0.3 |
| 8 | 4 | 0.2 |
| 9 | 2 | 0.1 |
| 10 | 4 | 0.2 |
| 11 | 2 | 0.1 |
| 12 | 1 | 0.0 |
| 18 | 2 | 0.1 |
| Total | 2,341 | 100.0 |

Table 3 shows the distribution of the CCP and CCP_P across the publication years. The table focuses on the years 2000 to 2014 in which 2,321 CCP_P have been mentioned. Since only 20 CCP_P have been mentioned between 1981 and 1999, these years are not considered in the table. Table 2 compares the distribution of the annual number of CCP with the distribution of the annual number of CCP_P. In this study, the annual percentages of CCP are used as the expected values of which the annual percentages of CCP_P more or less differ (see the column "Difference between percentages"). A positive (negative) percentage means that the authors of the policy documents are more (less) interested in the CCP of a year than one can expect.

Table 3. Differences between the number of annual climate change papers published and the annual number of papers mentioned in policy documents (at least one time). Papers from grey marked publication years are more frequently mentioned in policy documents than can be expected. Publication years before 2000 are not considered.

| Publication year | Number of climate change papers (CCP) | Number of climate change papers with at least one policy mention (CCP_P) | Difference between percentages |
|---|---|---|---|

|      | absolute | in percent | absolute | in percent |       |
|------|----------|------------|----------|------------|-------|
| 2000 | 4,533    | 2.4        | 43       | 1.9        | -0.5  |
| 2001 | 4,889    | 2.6        | 39       | 1.7        | -0.9  |
| 2002 | 5,262    | 2.8        | 108      | 4.7        | 1.9   |
| 2003 | 5,983    | 3.2        | 142      | 6.1        | 3.0   |
| 2004 | 6,594    | 3.5        | 174      | 7.5        | 4.0   |
| 2005 | 7,409    | 3.9        | 236      | 10.2       | 6.3   |
| 2006 | 8,519    | 4.5        | 210      | 9.0        | 4.6   |
| 2007 | 10,259   | 5.4        | 167      | 7.2        | 1.8   |
| 2008 | 12,373   | 6.5        | 203      | 8.7        | 2.2   |
| 2009 | 14,060   | 7.4        | 191      | 8.2        | 0.8   |
| 2010 | 16,671   | 8.8        | 197      | 8.5        | -0.3  |
| 2011 | 19,059   | 10.0       | 230      | 9.9        | -0.1  |
| 2012 | 21,849   | 11.5       | 184      | 7.9        | -3.6  |
| 2013 | 25,320   | 13.3       | 161      | 6.9        | -6.4  |
| 2014 | 27,016   | 14.2       | 36       | 1.6        | -12.7 |
| Total| 189,796  | 100.0      | 2,321    | 100.0      |       |

The time-curve of the citations referenced by scientific papers usually shows a distinct peak between two and four years after publication of the cited paper. However, many highly cited papers show the phenomenon of delayed recognition: their citation rate peaks many years after publication (Redner, 2005). As the results in Table 3 show it also needs more time to be mentioned in a policy document, because the period of origin of the policy documents is much longer than that of a typical scientific paper (see e.g. the *IPCC Reports* which appear only every 6-7 years, whereas a typical scientific paper arises within one year). This corresponds to the maximum of the difference between the percentages around 2005-2006 in the right column of Table 3.

Table 4. Number of climate change papers and number of papers mentioned in policy documents (at least once) broken down by journals (sorted by number of papers mentioned in policy documents). Only those journals are shown with at least 10 papers mentioned in policy documents. Papers of grey marked journals are more frequently mentioned in policy documents than can be expected.

| Journal | Number of climate change papers (CCP) | Number of climate change papers with at least one policy mention | Difference between percentages |
|---------|---------------------------------------|------------------------------------------------------------------|-------------------------------|

| | absolute | in percent | (CCP_P) absolute | in percent | |
|---|---|---|---|---|---|
| *Geophysical Research Letters* | 4,399 | 9.1 | 273 | 19.3 | 10.2 |
| *Journal of Geophysical Research* | 7,354 | 15.3 | 212 | 15.0 | -0.3 |
| *Climatic Change* | 2,929 | 6.1 | 100 | 7.1 | 1.0 |
| *Science* | 1,028 | 2.1 | 89 | 6.3 | 4.2 |
| *Global Environmental Change* | 767 | 1.6 | 85 | 6.0 | 4.4 |
| *Nature* | 1,171 | 2.4 | 62 | 4.4 | 2.0 |
| *Proceedings of the National Academy of Sciences of the United States of America* | 1,325 | 2.7 | 54 | 3.8 | 1.1 |
| *Climate Dynamics* | 2,454 | 5.1 | 51 | 3.6 | -1.5 |
| *Global Biogeochemical Cycles* | 616 | 1.3 | 37 | 2.6 | 1.3 |
| *Energy Policy* | 1,807 | 3.7 | 35 | 2.5 | -1.3 |
| *Global Change Biology* | 2,223 | 4.6 | 31 | 2.2 | -2.4 |
| *PLoS ONE* | 2,112 | 4.4 | 31 | 2.2 | -2.2 |
| *Journal of climate* | 5,036 | 10.4 | 27 | 1.9 | -8.5 |
| *Philosophical Transactions of the Royal Society of London, Series B - Biological Sciences* | 386 | 0.8 | 21 | 1.5 | 0.7 |
| *Philosophical Transactions: Mathematical, Physical and Engineering Sciences* | 405 | 0.8 | 21 | 1.5 | 0.6 |
| *Nature Climate Change* | 409 | 0.8 | 21 | 1.5 | 0.6 |
| *Bulletin of the American Meteorological Society* | 636 | 1.3 | 20 | 1.4 | 0.1 |
| *Climate Research* | 1,040 | 2.2 | 20 | 1.4 | -0.7 |
| *Ecological Economics* | 485 | 1.0 | 20 | 1.4 | 0.4 |
| *Water Resources Research* | 961 | 2.0 | 20 | 1.4 | -0.6 |
| *Environmental Research Letters* | 894 | 1.9 | 19 | 1.3 | -0.5 |
| *Wiley Interdisciplinary Reviews : Climate Change* | 259 | 0.5 | 16 | 1.1 | 0.6 |
| *International Journal of Climatology* | 2,205 | 4.6 | 16 | 1.1 | -3.4 |
| *Agriculture, Ecosystems and Environment* | 667 | 1.4 | 15 | 1.1 | -0.3 |
| *Nature Geoscience* | 346 | 0.7 | 15 | 1.1 | 0.3 |
| *Journal of Hydrology* | 1,375 | 2.9 | 15 | 1.1 | -1.8 |
| *Environmental Health Perspectives - National Institute of Environmental Health Sciences* | 156 | 0.3 | 13 | 0.9 | 0.6 |
| *Risk Analysis* | 133 | 0.3 | 12 | 0.8 | 0.6 |
| *Freshwater Biology* | 311 | 0.6 | 11 | 0.8 | 0.1 |
| *Science of The Total Environment* | 797 | 1.7 | 11 | 0.8 | -0.9 |
| *Hydrological Processes* | 993 | 2.1 | 11 | 0.8 | -1.3 |
| *Global and Planetary Change* | 1,261 | 2.6 | 11 | 0.8 | -1.8 |
| *The Lancet* | 48 | 0.1 | 10 | 0.7 | 0.6 |
| *Forest Ecology and Management* | 1,221 | 2.5 | 10 | 0.7 | -1.8 |

| Total | | 48,209 | 100.0 | 1415 | 100.0 | |

Table 4 shows the number of CCP and number of CCP_P broken down by the journals in which the papers had appeared. Not surprisingly, the prominent journals *Nature* and *Science* are comparatively often mentioned by policy documents. Their multidisciplinary orientation and the more general character of many papers published by them is a decisive advantage for being considered and mentioned comparatively frequently. In contrast, the journal *Geophysical Research Letters* at the top of Table 4 is far more subject-specific (concerning climate change). The journal publishes many more policy-relevant papers in general (19.3% among CCP_P) than one can expect (9.1% among CCP). This corresponds to the publication output in terms of the total number of climate change related papers published and the large proportion of highly cited papers (Haunschild et al., conditionally accepted for publication). Both facts may increase the probability of being mentioned in a policy document, too.

Table 5. Number of climate change papers and number of papers mentioned in policy documents (at least once) broken down by document type (sorted by number of papers mentioned in policy documents). Papers with grey marked document types are more frequently mentioned in policy documents than can be expected.

| Document type | Number of climate change papers (CCP) | | Number of climate change papers with at least one policy mention (CCP_P) | | Difference between percentages |
|---|---|---|---|---|---|
| | absolute | in percent | absolute | in percent | |
| Article | 209,837 | 94.5 | 2,057 | 87.9 | -6.6 |
| Review | 12,223 | 5.5 | 284 | 12.1 | 6.6 |
| Total | 222,060 | 100.0 | 2,341 | 100.0 | |

In agreement with the relatively high percentages of *Nature* and *Science* papers (i.e. papers with a more general character and thereby review-like) among CCP_P shown in Table 4, review papers are mentioned comparatively more frequently in CCP_P (12.1%) than one can expect from the distribution among CCP (5.5%) – as the results in Table 5 reveal. The opposite is true for research articles (87.9% versus 94.5%). Reviews summarize research results of a topic spread out in a variety of specific papers and are therefore more useful for the authors of policy documents. Possibly, the authors (and potential readers) of such documents are unable to cope with more specific research results and to assess their importance and significance for the different questions around climate change.

Table 6. Number of climate change papers and number of papers mentioned in policy documents (at least once) broken down by OECD subject categories (sorted by number of papers mentioned in policy documents). Papers from grey marked subject categories are more frequently mentioned in policy documents than can be expected.

| Subject category | Number of climate change papers (CCP) | | Number of climate change papers with at least one policy mention (CCP_P) | | Difference between percentages |
|---|---|---|---|---|---|
| | absolute | in percent | absolute | in percent | |
| Earth and related environmental sciences | 120,433 | 41.4 | 1,756 | 47.1 | 5.7 |
| Biological sciences | 48,633 | 16.7 | 427 | 11.5 | -5.3 |
| Social and economic geography | 12,281 | 4.2 | 396 | 10.6 | 6.4 |
| Other natural sciences | 9,314 | 3.2 | 269 | 7.2 | 4.0 |
| Agriculture, forestry, and fisheries | 20,386 | 7.0 | 166 | 4.5 | -2.6 |
| Economics and business | 5,691 | 2.0 | 134 | 3.6 | 1.6 |
| Health sciences | 4,498 | 1.5 | 120 | 3.2 | 1.7 |
| Environmental engineering | 18,076 | 6.2 | 96 | 2.6 | -3.6 |
| Civil engineering | 6,254 | 2.1 | 46 | 1.2 | -0.9 |
| Basic medicine | 3,191 | 1.1 | 42 | 1.1 | 0.0 |
| Political Science | 2,370 | 0.8 | 36 | 1.0 | 0.2 |
| Other agricultural sciences | 2,623 | 0.9 | 34 | 0.9 | 0.0 |
| Sociology | 2,767 | 1.0 | 32 | 0.9 | -0.1 |
| Unmatched Subject Codes | 1,200 | 0.4 | 29 | 0.8 | 0.4 |
| Veterinary science | 1,229 | 0.4 | 25 | 0.7 | 0.2 |
| Clinical medicine | 2,050 | 0.7 | 19 | 0.5 | -0.2 |
| Environmental biotechnology | 1,669 | 0.6 | 14 | 0.4 | -0.2 |
| Animal and dairy science | 1,229 | 0.4 | 13 | 0.3 | -0.1 |
| Mathematics | 1,336 | 0.5 | 12 | 0.3 | -0.1 |
| Physical sciences | 4,228 | 1.5 | 10 | 0.3 | -1.2 |
| Psychology | 2,746 | 0.9 | 9 | 0.2 | -0.7 |
| Other engineering and technologies | 2,917 | 1.0 | 7 | 0.2 | -0.8 |

| Chemical sciences | 3,147 | 1.1 | 7 | 0.2 | -0.9 |
| Mechanical engineering | 3,067 | 1.1 | 7 | 0.2 | -0.9 |
| Law | 753 | 0.3 | 4 | 0.1 | -0.2 |
| History and archaeology | 1,408 | 0.5 | 4 | 0.1 | -0.4 |
| Chemical engineering | 2,641 | 0.9 | 4 | 0.1 | -0.8 |
| Electrical engineering, electronic engineering, information engineering | 1,439 | 0.5 | 3 | 0.1 | -0.4 |
| Computer and information sciences | 1,232 | 0.4 | 3 | 0.1 | -0.3 |
| Philosophy, ethics and religion | 523 | 0.2 | 2 | 0.1 | -0.1 |
| Other social sciences | 450 | 0.2 | 1 | 0.0 | -0.1 |
| Educational sciences | 1,186 | 0.4 | 1 | 0.0 | -0.4 |
| Languages and literature | 146 | 0.1 | 1 | 0.0 | 0.0 |
| Total | 291,113 | 100.0 | 3,729 | 100.0 | |

Note. Since papers can be assigned to more than one subject category, many papers are multiply considered.

Based on the journals in which the CCP have been published the CCP (and the CCP_P) were assigned to broader subject categories as defined by the OECD (Paris, France).[3] Table 6 shows the number of CCP and number of CPP_P broken down by these subject categories. Climate change research is hardly comparable with classical research topics (for example like photovoltaics or even nanoscience as a broader field). Due to the fact that nearly all systems on this planet are affected by the impacts of a changing climate, a multitude of quite different disciplines all over science is interacting and cooperating in this demanding field of science. Therefore, it is reasonable to differentiate the impact of climate change research on policy documents with regard to the various disciplines and sub-disciplines involved. Their influence on policy documents can be assumed to be quite different.

The overrepresentation of the "Earth and related environmental sciences" papers among the CCP_P (see the positive difference between the CPP and CPP_P percentages at the top of Table 6) corresponds to the overrepresentation of the associated journals of this category (in particular the *Geophysical Research Letters*) presented in Table 4. The large positive difference between the CPP and CPP_P percentages for the category *Social and economic geography* may initially be surprising, because the category is far outside the "classical" climate change related categories, like earth sciences. However, in view of the rapid growth of the publication output dealing with adaptation, impacts and vulnerability of climate change and the emergence of related title words within climate change literature (see the results in Haunschild et al., conditionally accepted for publication), the overrepresentation of *Social and economic geography* is understandable.

---

[3] See http://ipscience-help.thomsonreuters.com/incitesLive/globalComparisonsGroup/globalComparisons/subjAreaSchemesGroup/oecd.html

### 4.2 Climate change papers mentioned in policy documents most frequently

As Table 7 shows, papers published in *Nature* and *Science* benefit from these prestigious and most visible journals and the more general character of their papers: Five of the nine CCP_P with the most mentions (at least ten) in policy documents have been published in these journals. The topics of the papers most frequently mentioned in policy documents are more notable: Eight out of nine papers are about adaptation and vulnerability of climate change and deal with agriculture (crop production, food security) and fishery (one paper discusses emission scenarios of greenhouse gases). Thus, the papers have the more practical consequences of climate change for society as a topic. Since the results of Haunschild et al. (conditionally accepted for publication) show that the share of the publication output of climate change papers assigned to agriculture is rather low (about 9% of the overall climate change literature), policy documents seem to focus on a part of the climate change literature which is especially interesting in the political context – the welfare of the population.

Table 7. Nine climate change papers with the most mentions (at least ten) in policy documents

| Paper | Number of policy |
|---|---|

| | mentions |
|---|---|
| Foley, J. A., Ramankutty, N., Brauman, K. A., Cassidy, E. S., Gerber, J. S., Johnston, M., . . . Zaks, D. P. M. (2011). Solutions for a cultivated planet. *Nature*, 478(7369), 337-342. | 18 |
| Lobell, D. B., Burke, M. B., Tebaldi, C., Mastrandrea, M. D., Falcon, W. P., & Naylor, R. L. (2008). Prioritizing Climate Change Adaptation Needs for Food Security in 2030. *Science*, 319(5863), 607-610. | 18 |
| Lobell, D. B., Schlenker, W., & Costa-Roberts, J. (2011). Climate Trends and Global Crop Production Since 1980. *Science*, 333(6042), 616-620 | 12 |
| Bharucha, Z., & Pretty, J. (2010). The roles and values of wild foods in agricultural systems. *Philosophical Transactions of the Royal Society of London B: Biological Sciences*, 365(1554), 2913-2926. | 11 |
| Lal, R. (2004). Soil Carbon Sequestration Impacts on Global Climate Change and Food Security. *Science*, 304(5677), 1623-1627. | 11 |
| Foley, J. A., DeFries, R., Asner, G. P., Barford, C., Bonan, G., Carpenter, S. R., . . . Snyder, P. K. (2005). Global Consequences of Land Use. *Science*, 309(5734), 570-574. | 10 |
| Allison, E. H., Perry, A. L., Badjeck, M.-C., Neil Adger, W., Brown, K., Conway, D., . . . Dulvy, N. K. (2009). Vulnerability of national economies to the impacts of climate change on fisheries. *Fish and Fisheries*, 10(2), 173-196. | 10 |
| Welcomme, R. L., Cowx, I. G., Coates, D., Béné, C., Funge-Smith, S., Halls, A., & Lorenzen, K. (2010). Inland capture fisheries. *Philosophical Transactions of the Royal Society of London B: Biological Sciences*, 365(1554), 2881-2896. | 10 |
| Anderson, K., & Bows, A. (2010). Beyond 'dangerous' climate change: emission scenarios for a new world. *Philosophical Transactions of the Royal Society of London A: Mathematical, Physical and Engineering Sciences*, 369(1934), 20-44. | 10 |

# 5    Discussion

In recent years, societal impact measurements of academic research have become more and more important. This trend is not only visible by their consideration in national evaluation systems (e.g. the REF), but also in the commercial success of providers delivering altmetrics data (e.g. Altmetric) which propose that altmetric scores can be used to measure societal impact. Currently, the most important and most frequently used method of societal impact measurement is the case study approach in which cases of research are described leading successfully to a specific form of societal impact (King's College London and Digital Science, 2015). However, case studies have the disadvantages that they are expensive, the results are biased towards success stories, and the results for different entities (e.g. universities) are not comparable. Based on the results of case studies, it is not possible to say that one entity is more successful in generating societal impact than another entity. Such comparisons are only possible by using quantitative indicators with a large coverage of research entities: "The advantage of using quantitative indicators is that they can be standardized and aggregated, allowing universities to use them on a continuous basis to track their impact, compare it with other universities, and recognise the contribution of every faculty member, of whatever scale" (Ovseiko et al., 2012).

Whereas bibliometric indicators have emerged as the most important metrics to measure the recursive impact of research, the development of metrics for the measurement of societal impact is challenging. According to the US National Research Council (2014) "no high-quality metrics for measuring societal impact currently exist that are adequate for evaluating the impacts of federally funded research on a national scale … Each metric describes but a part of the larger picture, and even collectively, they fail to reveal the larger picture. Moreover, few if any metrics can accurately measure important intangibles, such as the knowledge generated by research and research training" (p. 70). In section 2, we refer to Wilsdon et al. (2015) who highlight only two quantitative indicators which can be used for the societal impact measurement: Google patent citations (for measuring innovation in industry) and clinical guideline citations (for measuring the impact of biomedical research on clinical practice). However, further indicators are necessary which allow targeting the three institutional foundations of impact: (i) epistemological (better understanding of phenomena behind different kinds of societal problems), (ii) artefactual (development of technological artefacts and instruments), and (iii) interactional (organizational forms of partnerships between researchers and different kinds of societal actors) (Miettinen et al., 2015).

This study focusses on a relatively new form of impact measurements (provided by Altmetric), which could complement Google patent citations and clinical guideline citations: mentions of publications in policy documents. It is an interesting form of impact measurement compared to other altmetrics (e.g. mentions in tweets and blogs) because (1) it is target-oriented (i.e. it measures the impact on a specific sector of society) and (2) it focusses on a relevant part of society for research – the policy area. Many research topics are policy-relevant (e.g. health care or labor market research) and it is interesting to know in the context of wider impact evaluations which (kind of) publications have more or less impact. Altmetric and Scholastica (2015) exemplify that policy document mentions cannot only be used on the institutional level to demonstrate impact, but also on the level of single researchers. They describe the case of a university professor who would like to show the broader impact of her research to the National Institutes of Health (Bethesda, MD; USA) and the Medical Research Council's (London UK) program officers: "Altmetrics were able to show her that her work had been referenced in policy documents published by two major organizations – evidence she considered 'bona fide data demonstrating that practitioners – not researchers – but folks who can affect lives through legislation, health care, and education, are using my research to better their work'" (p. 26).

In this study, we used a comprehensive dataset of papers on climate change to investigate mentions of papers in policy documents. Climate change is particularly useful in this respect because the topic is very policy relevant since many years. Thus, we expected to find a large number of papers mentioned in policy documents in comparison with other research fields – especially because corresponding policy sites are continuously evaluated by Altmetric. However, the results of the analyses could not validate our expectation: Of n=191,276 publications on climate change in the dataset, only 1.2% (n=2,341) have at least one policy mention. The rate of 1.2% is also small in comparison with the result of Kousha and Thelwall (in press) who show that "within Biomedical Engineering, Biotechnology, and Pharmacology & Pharmaceutics, 7% to 10% of Scopus articles had at least one patent citation". The result of this study contradicts the claim of Khazragui and Hudson (2015) that "it is rare that a single piece of research has a decisive influence on policy. Rather policy tends to be based upon a large body of work constituting 'the commons'" (p. 55). The low percentage of 1.2% which we found in this study might be due to the fact that Altmetric quite recently started to analyze policy documents and the coverage of the literature is still low (but will be extended). However, the low percentage might also reflect that only a small part of the literature is really policy relevant and most of the papers are only relevant for researchers studying climate change. Two other reasons for the low percentage might be that (1) policy documents may not mention every important paper on which a policy document is based on. (2) There are possible barriers and low interaction levels between researchers and policy makers.

The low number of mentions in policy documents further raises the question what mentions in policy-related documents really measure: Is it relevance of academic papers? Do the mentions reflect the effort of researchers to interact with policy makers, an ongoing relationship between researchers and policy makers, or the effort of the policy organization to include (climate change) research in policy documents? Future studies should try to find an answer on these questions by undertaking (1) analyses of the context of paper mentions in policy-related documents or (2) surveys of the authors of these documents asking for the motivations for the paper mentions. Independent of these and other possible results of forthcoming studies, one should keep in mind that non-mentioned papers are not necessarily less relevant than mentioned papers; they may simply be unknown to policy makers.

In order to find out which kind of papers are more or less interesting in the policy context (e.g. articles or reviews), we compared the distribution of papers among CCP and CCP_P. The results show that the policy literature tends to cite research which has been published a longer time ago than researchers do in their papers. Thus, research papers seem to need more time to produce impact on politics than on research itself. As expected, reviews are overrepresented among CCP_P: the observed CCP_P value is higher than the expected value delivered by the CCP distribution. Reviews summarize the results of many primary research papers and connect research lines from different research groups. Good reviews save the labor of reviewing the literature on one's own responsibility. In this study, we further revealed that papers published in *Nature* and *Science* as well from the areas "Earth and related environmental sciences" and "Social and economic geography" are especially relevant in the policy context.

This study is a first attempt to study mentions of scientific publications in policy documents. We encourage that further empirical studies follow because the data source is of special interest in the use of altmetric data for measuring the broader impact of research. It will be interesting to see whether more papers are used in policy documents in upcoming years (because of the wider coverage of the policy literature by Altmetric). Furthermore, it would be interesting to generate results from other areas of research (which are also particularly interesting for the policy area) in order to compare the results for climate change with the results for other areas. For example, it will be interesting to see whether there are higher or lower percentages than 1.2% of publications mentioned in policy documents (see above). Do policy documents from other areas focus on more recent literature than policy documents on climate change do? When the coverage of research papers in policy documents will be on a significantly higher level, future studies should also try to normalize policy document mentions. This study has already demonstrated that mentions in policy documents are time-dependent and we can expect that policy documents will differently focus on certain areas of research. Thus, time- and area-specific forms of normalization will be necessary if entities (e.g. research groups or institutions) with broad outputs in times and areas are evaluated.

In conclusion, we would like to mention limitations of this study. The first one refers to the documents analyzed by Altmetric. Altmetric does not only analyze documents from governments, but also documents from researchers who summarize the status of research on a certain topic for politicians as potential readers. Bornmann and Marx (2014) proposed that the former document type can be named as assessment report: Assessment reports summarize the status of research on a certain subject in form of narrative reviews or meta-analyses. However, assessment reports are preliminary stages of impact on politics, because they summarize the literature with the goal of increasing the impact of research on politics. Real impact on politics can only be measured by analyzing documents from governments. A second limitation of this study is the problem of disambiguating policy documents in different and same languages. Some policy documents have the same content while only the language differs. In these cases, the mentions are counted multiple times.

## 6   Conclusions

Bibliometrics is particularly successful in measuring impact, because the target of impact measurement is clearly defined: the publishing researcher who is working in the science system. Thus, citation counts provide target-oriented metrics (Lähteenmäki-Smith, Hyytinen, Kutinlahti, & Konttinen, 2006). However, many societal impact measurement studies are intended to measure impact in a broad sense whereas broad means the impact on all areas of society (or at least as many as possible). This is especially the case for studies which are based on various sources of altmetrics. Furthermore, there is the tendency in altmetrics to use different altmetric sources for the calculation of a composite indicator. Composite indicators are calculated because many altmetric sources are characterized by low counts. For example, Altmetric has proposed the Altmetric Attention Score, which summarizes the impact of a piece of research over different altmetric sources (e.g. blog posts and tweets) (https://help.altmetric.com/support/solutions/articles/6000059309-about-altmetric-and-the-altmetric-attention-score).

We deem appropriate that the measurement of impact should always be target-oriented. Altmetric and Scholastica (2015) give some examples here: "For example, someone publishing a study on water use in Africa may be particularly keen to see that many of those tweeting and sharing the work are based in that region, whereas economics scholars might want to keep track of where their work is being referenced in public policy or by leading think-tanks" (p. 24). Thus, we appreciate those quantitative approaches which analyses scholarly paper citations, clinical guideline citations, policy document mentions, or patent citations. Without the restriction of impact to specific target groups (identifiable recipients) it is not clear what kind of impact is actually measured. The general uncertainty in scientometrics about the meaning of altmetrics is probably based on the tendency to use composite indicators or counts without target-restrictions when altmetrics are applied.

Policy documents are one of the few altmetric sources which can be used for the target-oriented impact measurement. As source for the measurement of impact on politics, policy documents are one of the most interesting altmetric sources which should be studied in more detail in future studies. For the use of a metric based on policy document mentions in these studies, we deem it as very necessary that Altmetric publishes an up-to-date list of the sources with policy-related documents which they continuously analyze. Without this information it is scarcely possible to interpret the results of a study which is based on mentions in policy-related documents. On the current homepage, Altmetric only lists some examples of the type of organizations included in the database (Liu, 2014). Many national sources of policy documents, such as ministries or regional governments, are not mentioned in the list. Are these sources not included in the Altmetric database? We suggest to list in detail the types of policy making bodies included in the database and the number of organizations per type (perhaps with examples). This gives the user an idea of the inclusiveness of the database and thereby allows them to make a more informed assessment of the results of the analysis. If the typology would also contain the total number of papers mentioned, it could show which type of policy making bodies already cooperate closely with researchers.

Taken as a whole, many questions should be answered until one can decide whether this new altmetric source can be used in practice, i.e. in the evaluation practice of the REF,

ERA or others. Currently, there are too many open questions which disregard the new source as a metric for a prompt consideration. The user of the new data source and recipient of corresponding results should always keep in mind that the analysis is of quantitative nature; counting the number of mentions of a paper set in policy documents. It is not a qualitative analysis of how the research described in the paper set is being discussed in policy documents. This information could only be retrieved when the context of mentions in policy documents is analyzed (Bornmann & Daniel, 2008). Furthermore, the analysis of mentions in policy-related documents cannot uncover the different forms of interactions between science and policy. In section 2, we described two typologies which have been developed to describe these forms. The quantitative analysis of mentions in policy-related documents treats the mentioned scientist as a *pure scientist* who is oriented towards facts and is actually not interested on the interaction with political actors (Pielke, 2007). The interaction is reduced to the reading of papers (which are oriented to academic audiences) by political actors who deemed the papers as so important that they mention them in the policy-related document.

# Acknowledgements

The bibliometric data used in this paper are from CrossRef and an in-house database developed and maintained by the Max Planck Digital Library (MPDL, Munich) and derived from the Science Citation Index Expanded (SCI-E), Social Sciences Citation Index (SSCI), Arts and Humanities Citation Index (AHCI) prepared by Thomson Reuters (Philadelphia, Pennsylvania, USA). The policy document information was retried via the API of Altmetric.